\documentclass[final,5p,times,twocolumn,authoryear]{elsarticle}
 
\usepackage{amssymb}
\usepackage{lipsum}\usepackage{amsmath}
\journal{Physics Letters B}
\usepackage[format=hang,font=small,labelfont=bf]{caption}
\usepackage[colorlinks=true,citecolor=blue,linkcolor=red,urlcolor=blue]{hyperref}

\begin{document}

\title{Non-linear equation of motion for higher curvature semiclassical gravity}
\author{Naman Kumar}
\date{}
\affiliation{organization={Department of Physics, Indian Institute of Technology Gandhinagar, Palaj, Gujarat, India, 382355},
email={namankumar5954@gmail.com}}

\begin{abstract}
We derive the non-linear semiclassical equation of motion for a general diffeomorphism-invariant theory of gravity by leveraging the thermodynamic properties of closed causal horizons. Our work employs two complementary approaches. The first approach utilizes perturbative quantum gravity applied to a Rindler horizon. The result is then mapped to a stretched light cone, which can be understood as a union of Rindler planes. Here, we adopt the semiclassical physical process formulation, encapsulated by $\langle Q\rangle = T \delta S_{gen}$ where the heat-flux $\langle Q\rangle$ is related to the expectation value of stress-energy tensor $T_{ab}$ and $S_{gen}$ is the generalized entropy. The second approach introduces a "higher curvature" Raychaudhuri equation, where the vanishing of the quantum expansion \(\Theta\) pointwise as required by restricted quantum focusing establishes an equilibrium condition, \(\delta S_{\text{gen}} = 0\), at the null boundary of a causal diamond. While previous studies have only derived the linearized semiclassical equation of motion for higher curvature gravity, our work resolves this limitation by providing a fully non-linear formulation without invoking holography. 
\\ \paragraph*{Keywords:}Higher curvature gravity; Generalized entropy; Entanglement equilibrium; Perturbative Quantum gravity; Stretched light cone; Causal diamond.
\end{abstract}
\maketitle

\newpage
\section{Introduction}
The connection between thermodynamics and black holes was established by the works of Bekenstein and Hawking \cite{Bekenstein:1972tm,bekenstein1973black,hawking1975particle}. Jacobson \cite{jacobson1995thermodynamics} advanced a viewpoint by deriving Einstein's equation using local equilibrium conditions and Clausius entropy relation $Q=T\delta S$ where $Q$ is the heat flux given as the boost Killing energy, $T$ is the Unruh temperature, and $S=A/4G$ is the horizon entropy. It was later extended to the non-equilibrium case by Eling et al. \cite{eling2006nonequilibrium} where they derived $f(R)$ theory of gravity. In 2016, Jacobson \cite{jacobson2016entanglement} refined his original arguments by deriving (semiclassical) Einstein's equation using the entanglement equilibrium approach where he showed that for a causal diamond, if matter field and geometry are varied simultaneously, then the entanglement entropy vanishes at first order\footnote{We set $\hbar=1$ in this paper.}
\begin{equation}
    \delta S_{EE}|_V=\frac{\delta A|_V}{4G}+\delta S_{IR}=0.\label{ee_variation}
\end{equation}
where $S_{IR}$ is the (finite) IR part of entanglement entropy. Jacobson assumed that the theory of quantum gravity is UV finite (this is an assumption since we know nothing about the theory of quantum gravity) since it is known that entanglement entropy $S_{EE}$ for a $d+1$ quantum field theory (QFT) is UV divergent and is given as
\begin{equation}
    S_{EE}=c_0\frac{A}{\epsilon^{d-1}}+S_{finite}+\text{subleading divergence}.
\end{equation}
Subleading divergence arises from higher curvature corrections as required by loop quantum calculations or stringy physics. If we identify $\frac{c_0}{\epsilon^{d-1}}\to\frac{1}{4G}$, we get Bekenstein-Hawking entropy as leading divergence. Then, the entanglement entropy is identified as the generalized entropy $S_{gen}$
\begin{equation}
    S_{EE}=S_{gen}=S_{grav}+S_{matter},
\end{equation}
which is known to be a cutoff independent quantity \cite{susskind1994black}. Here, $S_{grav}=A/4G$ if the theory of gravity is governed by Einstein-Hilbert action and is equal to Wald entropy $S_{Wald}$ if the theory is described by a higher curvature action. Although it is tempting to identify the entanglement entropy of quantum states outside the horizon with generalized entropy, in our understanding, the dependence of entanglement entropy on a (fine-tuned) UV cut-off makes it challenging to do so.\\
For small diamonds, (\ref{ee_variation}) leads to the non-linear semiclassical Einstein's equation. The (linearized) semiclassical equation of motion was earlier derived using AdS/CFT correspondence in \cite{lashkari2014gravitational,faulkner2014gravitation,swingle2014universality}. Jacobson's entanglement equilibrium derivation was extended to any general diffeomorphism invariant theory of gravity using a causal diamond in \cite{bueno2017entanglement} and using a stretched light cone in \cite{svesko2019entanglement}. \\
To remind the readers, a general theory of gravity\footnote{From now on, we shall refer to a general diffeomorphism invariant theory of gravity as the general theory of gravity or higher curvature gravity interchangeably.} is one in which Lagrangian depends on both metric tensor and Riemann curvature tensor and is expressed as
\begin{equation}
    I=\frac{1}{16\pi G}\int d^4x\sqrt{-g}L(g_{ab},R_{abcd})+I_{matter}.
\end{equation}
Here $I_{matter}$ is the action of the matter field. If we define
\begin{equation}
    P^{abcd}=\frac{\partial L}{\partial R_{abcd}},
\end{equation}
we get the equation of motion as
\begin{equation}
    P^{cde}_aR_{bcde}-2\nabla^c\nabla^dP_{abcd}-\frac{1}{2}Lg_{ab}=8\pi G T_{ab}.
\end{equation}
Both the derivations concluded that for higher curvature gravity, it is not possible to derive the full non-linear semiclassical equation of motion using the entanglement equilibrium approach. For General Relativity (GR), the higher order expansion in Riemann normal coordinates (RNC) for a geodesic ball goes as $l^2/L^2$ where $l$ is the size of the diamond, and $L$ is the curvature scale. Suppose we consider the diamond to be much smaller than the curvature scale. In that case, the higher-order terms are already suppressed, and the linearized equations obtained by perturbation around the Minkowski diamond are equivalent to the full non-linear equation of motion. However, in the case of higher curvature gravity, higher order terms in the RNC expansion are needed to capture the non-linear terms in curvature, but these terms contribute in the same order as the non-linear terms; therefore, the perturbative expansion breaks down, and the conclusion that linearized equation implies the full non-linear tensorial structure of field equation cannot be reached \cite{bueno2017entanglement}. \\In \cite{kumar2023recovering}, the non-linear semiclassical Einstein's equation was derived using a completely different approach utilizing quantum focusing conjecture \cite{bousso2016quantum} and generalized entropy. The difference in entanglement equilibrium and generalized entropy approach of \cite{kumar2023recovering} is a fundamental one: the former is a \textit{variational version} where two infinitesimally nearby geometries are considered while the latter proposes a semiclassical \textit{physical process version} as
\begin{equation}
    \langle Q\rangle=T\delta S_{gen}.\label{pp}
\end{equation}
The non-linear equation of motion in the classical case for a general theory of gravity has been derived in \cite{parikh2016generalized} using a stretched timelike surface of a Rindler horizon and in \cite{parikh2018einstein} using a stretched future light cone. In \cite{parikh2020local}, it was shown the local first law holds for small segments of a stretched light cone, thereby justifying the attribution of thermodynamic properties to stretched future light cones. The non-linear semiclassical equation of motion for higher curvature gravity has not yet been derived.\\
In this letter, we complete this piece of the puzzle by deriving non-linear semiclassical equation of motion for a general theory of gravity.\\
The paper is organized as follows: in section II, we derive the semiclassical equation of motion for a general theory of gravity using a stretched light cone. In section III, we employ a different method to derive the semiclassical equation of motion using a causal diamond. Physically, the difference in both approaches arises from the method of deriving the equation of motion. In the first approach, we use an approach similar to the first law of entanglement entropy, while the latter method is closer in spirit to Jacobson's entanglement equilibrium. Therefore, both approaches complement each other and make the paper more complete. We end the paper with some discussion.
\section{Stretched Light Cone}
\begin{figure}[ht!]
    \centering
    \includegraphics[width=\linewidth]{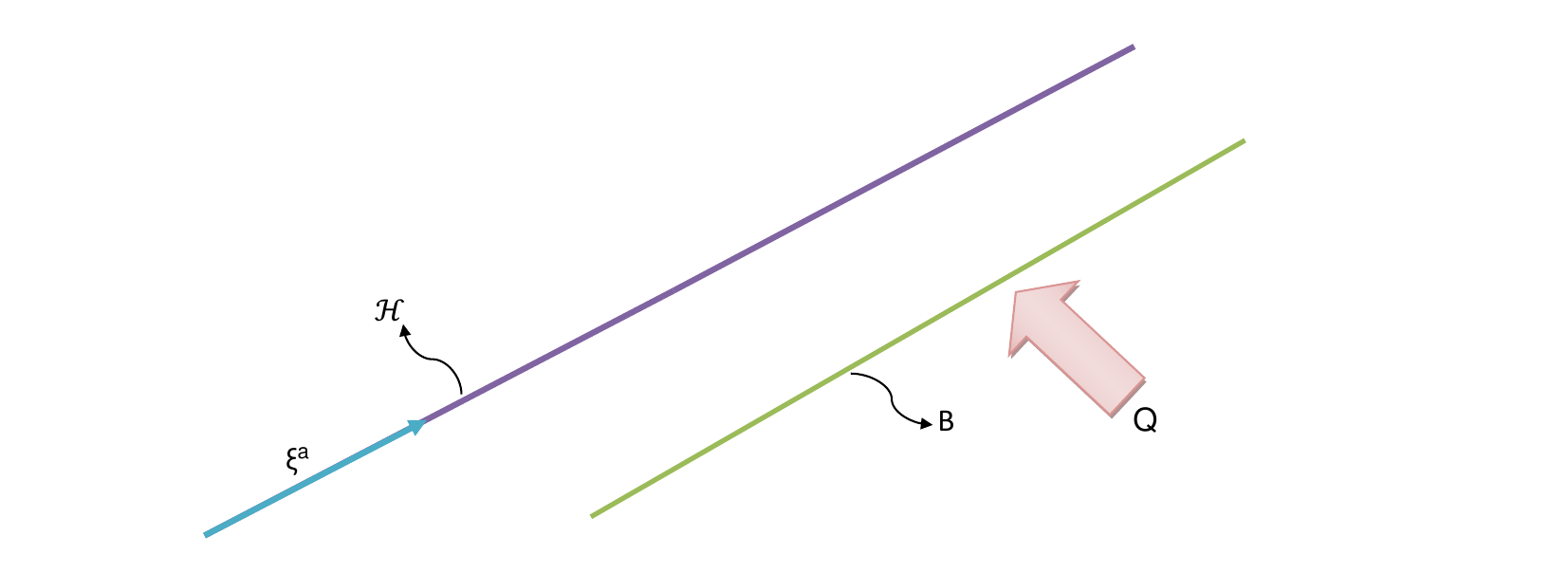}
    \caption{A Rindler horizon $\mathcal{H}$. $B$ is a codimension-2 surface spanned by light rays. When a matter flux $Q$ crosses the beam, it gets perturbed. The difference in the asymptotic area of the unperturbed and perturbed beam gives the change in the area of the Rindler horizon $\delta A$.}
    \label{light_surface}
\end{figure}

\begin{figure}[ht!]
    \centering
    \includegraphics[width=\linewidth]{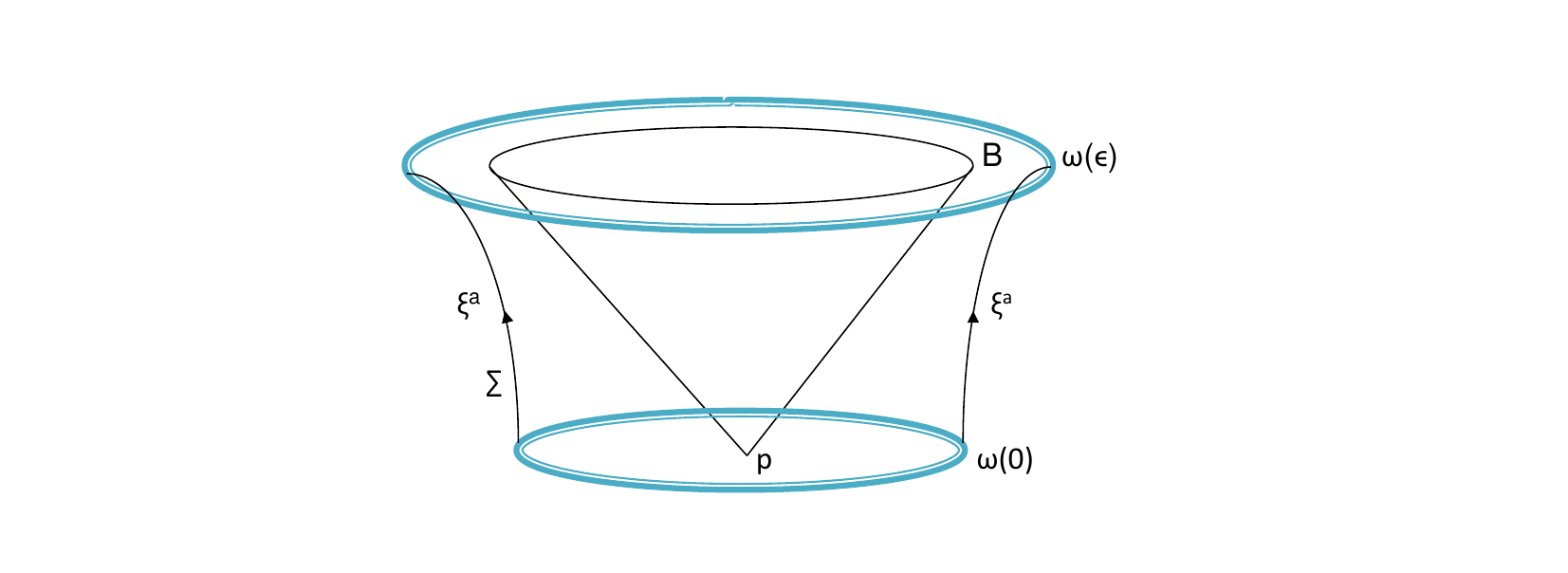}
    \caption{A stretched future light cone at a point $p$ is generated by a congruence of radially accelerating worldlines with the same proper acceleration $1/\alpha$. It describes a timelike hypersurface $\Sigma$. The boundary is given by two constant time slices of $\Sigma$ at time $t=0$ and $t=\epsilon$.}
    \label{stretched_lc}
\end{figure}

A thermodynamic law for the Rindler horizon relating the change in its entanglement entropy with the area was derived using perturbative quantum gravity in \cite{bianchi2013blackholeentropygraviton,bianchi2013mechanical}. Although the entanglement entropy has a UV divergent area term and requires a fine-tuned UV cutoff to be equal to generalized entropy, it was shown that the \textit{change} in entanglement entropy is finite under the assumptions that the energy of perturbation is less than some UV-cutoff ($E<E_{UV}$) and wavelength of perturbation is smaller than the curvature scale at the horizon ($l<l_{curv}$). These assumptions ensure that the change in entanglement entropy is insensitive to the divergent UV part and that all the contributions come from the finite IR part and one exactly reproduces the Bekenstein-Hawking entropy\footnote{Note that although we are in the semiclassical regime, the entanglement entropy by definition is an expectation value. Therefore, the area term on the right-hand side should also be calculated as the expectation value of the area operator (by quantizing (\ref{area_operator})). However, for the sake of brevity, we will simply use $\delta A$.}
\begin{equation}
    \frac{\langle Q\rangle}{T}=\delta S_{EE}=\frac{\delta \langle A\rangle}{4G}.\label{heat_area_quantum}
\end{equation}
Here, heat-flux is related to the expectation value of stress-energy tensor $T_{ab}$. This \textit{physical process} version bears resemblance with the first law of entanglement entropy $\frac{\delta\langle K\rangle}{T}=\delta S_{EE}$ via Ryu-Takayanagi prescription \cite{ryu2006holographic,ryu2006aspects} (where $K$ is the modular Hamiltonian) and is therefore capable of imposing the semiclassical equation of motion using thermodynamics rather than entanglement. An immediate advantage is that being a physical process version, it can be used to derive the non-linear equation of motion rather than only the linearized equation of motion, a key feature of variational version approaches.\\
There is another way to look at this interesting result: If the horizon entropy is identified as entanglement entropy, then the entanglement entropy of states outside the horizon is the same as the (fine-grained) generalized entropy \cite{sorkin1986toward,sorkin1997statisticalmechanicsblackhole,solodukhin2011entanglement}. Therefore, (\ref{pp}) takes the form
\begin{equation}
    \langle Q\rangle=T\delta S_{gen}=\frac{\delta A}{8\pi G}.
\end{equation}
The relation (\ref{heat_area_quantum}) is derived using perturbative gravity by considering a light beam (see Fig.(\ref{light_surface})) perturbed by gravity. In this case, we consider a small perturbation around Minkowski space as
\begin{equation}
    g_{\mu\nu}=\eta_{\mu\nu}+\sqrt{32\pi G}h_{\mu\nu}.
\end{equation}
At the lowest order in $G$, Einstein's equation becomes
\begin{equation}
    \Box h_{\mu\nu}=-\sqrt{8\pi G}(T_{\mu\nu}-\frac{1}{2}\eta_{\mu\nu}T^\alpha_\alpha).\label{eom}
\end{equation}
where $T_{\mu\nu}$ is the energy-momentum tensor of matter plus gravitational field 
\begin{equation}
    T_{\mu\nu}=T^{matter}_{\mu\nu}+T^{gravity}_{\mu\nu}.
\end{equation}
The change in asymptotic area between perturbed and unperturbed light beam is given as
\begin{equation}
    \delta A=-\int_{B_\infty} d^2y\int_0^\infty\dot\theta(v)vdv\label{area_variation},
\end{equation}
where $B_v$ denotes a two-dimensional surface corresponding to the unperturbed beam. It is coordinated by $\sigma^\alpha=(\sigma_1,\sigma_2)$, and its embedding in Minkowski spacetime is given in Cartesian coordinates by $x^\mu_0(v,\sigma)$. The corresponding surface of the perturbed beam is given by the embedding
\begin{equation}
    x^\mu(v,\sigma)=x^\mu_0(v,\sigma)+\xi^\mu(v,\sigma),
\end{equation}
and $B_\infty$ corresponds to the asymptotic area as $v\to\infty$. $\theta$ is the expansion defined as $v$-derivative of the perturbation of area density
\begin{equation}
    \theta=\partial_\alpha\dot\xi^\alpha+\sqrt{8\pi G}\dot h^\alpha_\alpha.
\end{equation}
(\ref{area_variation}) can be written as
\begin{equation}
    \delta A=\sqrt{8\pi G}\int d^2y\int_0^\infty dv v[-\Box h_{\mu\nu}(x_0(v))]k^\mu k^\nu.\label{area_operator}
\end{equation}
This is equal to the change in the area of the Rindler horizon since these null rays span a Rindler horizon.
Then using the equation of motion (\ref{eom}) and the expression 
\begin{equation}
\begin{split}
\delta S_{vN}&=-\mathrm{Tr}(\delta\rho\log\rho_0)=2\pi\mathrm{Tr}(K\delta\rho)\\
&=2\pi\int d^2y\int_0^\infty dv\hspace{1mm}\kappa\hspace{1mm} v\hspace{1mm} T_{\mu\nu}k^\mu k^\nu,  
\end{split}
\end{equation}
we finally obtain (\ref{heat_area_quantum}). Here, $K$ is the Rindler Hamiltonian. The relation (\ref{heat_area_quantum}) can equally be interpreted as a physical process version as is clear from (\ref{area_variation}) (the area variation also equals the expansion of the perturbed beam from $v=0$ to $v=\infty$). Therefore, for the setup we discussed, (\ref{pp}) takes the following form 
\begin{equation}
    \frac{\langle Q\rangle}{T}=\frac{\delta A}{4G}.
\end{equation}
Bianchi's result hinges on using the equation of motion (EoM) to relate the metric perturbation around the Minkowski background to the energy-momentum tensor under a perturbative expansion. For higher curvature gravity, the EoM (in a suitable gauge) under a perturbative expansion around the Minkowski background schematically take the form:
\begin{equation}
    \Box h_{ab}+\text{additional terms}=T_{ab}.
\end{equation}
In Bianchi's derivation, the first term directly gives rise to the area law ($\delta S\propto\delta A$) due to the standard relation between the metric perturbation and the horizon area. For higher curvature gravity, the additional terms modify this area law by introducing higher-order corrections. These corrections are captured by the Wald entropy $S_{Wald}$, which generalizes the Bekenstein-Hawking entropy to account for higher curvature contributions.
Therefore, in the context of higher curvature gravity, it is natural to extend Bianchi's result and consider 
\begin{equation}
    \frac{\langle Q\rangle}{T}=\delta S_{Wald},
\end{equation}
where $S_{Wald}$ encodes the entropy variation consistent with the modified area law.\\
Now that we are equipped with all the tools, our next task is to evaluate the Wald entropy. But before we evaluate the Wald entropy, we note that the above results were derived for a Rindler horizon, and therefore, these results directly map and hold for a stretched light cone \cite{parikh2018einstein} (see Fig.(\ref{stretched_lc})). Intuitively, a stretched light cone can be viewed as a union of Rindler planes, making it natural to expect that results derived for the Rindler horizon directly map to and hold for the stretched light cone. More precisely, the key ingredient in Bianchi’s derivation is the well-defined expression for the modular Hamiltonian. For a Rindler horizon, the modular Hamiltonian is proportional to the boost operator, which is well-known in this setup. Similarly, for a stretched light cone, the modular Hamiltonian is proportional to the radial boost operator \cite{svesko2019entanglement}, ensuring consistency between the two cases. Technically, the differences arise only in the expressions for the modular Hamiltonian $K$ and area difference $\delta A$. For a Rindler horizon, these are expressed in terms of the null parameter $v$ and the null vector $k^a$, which are specific to null surfaces. In contrast, for a stretched light cone, the modular Hamiltonian and area difference are expressed in terms of the normal vector $n^a$, proper time $\tau$, and velocity vector $u^a$, which are specific to the timelike surface $\Sigma$. Despite these adjustments, the underlying thermodynamic relationships remain unchanged. Thus, the results derived for the Rindler horizon extend naturally to the stretched light cone with only minor technical modifications.
\\ The reasons for using a stretched light cone are that: First, it is a closed causal horizon, and the notion of (entanglement) entropy is well-defined, while a finite patch of Rindler horizon unlike an infinite global Rindler horizon, does not separate regions into two parts, and the notion of entropy does not seem to be well-defined\footnote{This does not conflict with Jacobson's 1995 derivation since it was only assumed that the patch of horizon possesses entropy equal to its area and Bekenstein-Hawking entropy was not interpreted as an entanglement entropy.}. Second, a generic spacetime does not admit a true Killing vector but an approximate Killing vector and therefore, the Killing identity will fail in some order. Third, a Rindler horizon has boundaries, and when applying Stoke's theorem to calculate Wald entropy, we will have extra unwanted terms. Since a stretched light cone is spherical, it has no boundary, and we can evade this problem naturally. These technical issues were first addressed in \cite{guedens2012horizon}; however, some problematic assumptions were made, such as entropy is not equal to area, even for Einstein's theory of gravity. Therefore, working with the stretched light cone, we can write the Wald entropy at time $t$ as
\begin{equation}
    S_{Wald}=-\frac{1}{4G}\int_{\omega(t)} dS_{ab}(P^{abcd}\nabla_c\xi_d-2\xi_d\nabla_cP^{abcd}).
\end{equation}
Using Stoke's theorem for anti-symmetric tensor, the total change in Wald entropy between time $t=0$ and $t=\epsilon$ is
\begin{equation}
\begin{split}
    \delta S_{Wald}&=\frac{1}{4G}\int_\Sigma[-\nabla_b(P^{adbc}+P^{acbd})\nabla_c\xi_d\\
    &+P^{abcd}(R_{dcbe}\xi^e+f_{bcd})-2\xi_d\nabla_b\nabla_c P^{abcd}]d\Sigma^a.
\end{split}
\end{equation}
Since $\xi^a$ is not a true Killing vector in a generic spacetime, we use the technique outlined in \cite{parikh2018einstein} to discard unwanted terms $f_{bcd}$ arising from the failure of Killing identity to hold exactly and subtract the entropy due to the natural increase of hyperboloid (an irreversible process). This results in 
\begin{equation}
    \delta S_{rev}=\frac{1}{8\pi\alpha G}\int_\Sigma(P^{abcd}R_{dcbe}\xi^e-2\xi_d\nabla_b\nabla_c P^{abcd})d\Sigma^a.
\end{equation}
The heat flux across $\Sigma$ is
\begin{equation}
    \langle Q\rangle=\int_\Sigma d\Sigma_a\langle T^{ab}\rangle u^b=\frac{1}{\alpha}\int_\Sigma d\Sigma_a\langle T^{ab}\rangle \xi^b.
\end{equation}
Using the Clausius relation $\langle Q\rangle=T\delta S_{rev}$, we obtain
\begin{equation}
   P_a^{cde}R_{bcde}-2\nabla^c\nabla^dP_{acdb}+fg_{ab}=8\pi G\langle T_{ab}\rangle.
\end{equation}
where $f$ is some function. Imposing conservation on both sides, we finally obtain the full non-linear semiclassical equation of motion for higher curvature gravity as
\begin{equation}
     E_{ab}\equiv P_a^{cde}R_{bcde}-2\nabla^c\nabla^dP_{acdb}-\frac{L}{2}g_{ab}+\Lambda g_{ab}=8\pi G\langle T_{ab} \rangle.
\end{equation}
where $\Lambda$ is a constant identified as the cosmological constant. This identification is a usual practice in the derivation of the equations of motion using thermodynamics. Note that since the area in (\ref{heat_area_quantum}) is a quantum expectation value, the left-hand side of the above equation should also be treated as a quantum expectation value $\langle E_{ab}\rangle$ (see \cite{jacobson2016entanglement} for the same argument in the derivation of Einstein's equation using entanglement equilibrium). 
\section{Causal Diamond}
Consider a causal diamond for a ball-shaped region $\Sigma$ in a Maximally Symmetric Spacetime (MSS). A causal diamond in an MSS admits a conformal Killing horizon (CKH). A CKH has well-defined thermodynamic properties, and its shear vanishes \cite{dyer1979conformal}- a property that we make use of. This makes the diamond in MSS useful for our study.\\ For a closed causal horizon, we have a well-defined notion of entanglement entropy, which is given as the sum of a UV divergent part (which is regularized by choosing a cutoff) and a finite IR part or matter entanglement entropy. This makes them ideal to work in the semiclassical case where matter entanglement entropy becomes important. We work on the null boundary $\mathcal{H}$, which is the conformal Killing horizon \cite{dyer1979conformal}. 
\begin{figure}[ht!]
    \centering
    \includegraphics[width=\linewidth]{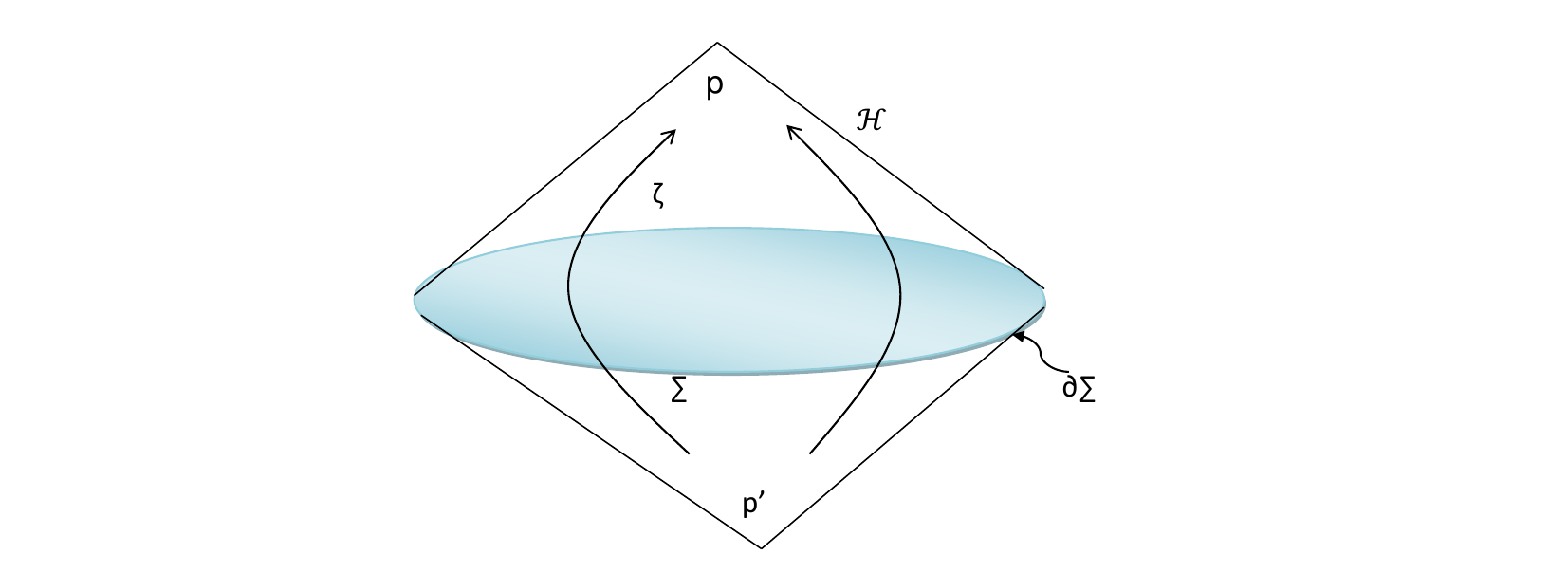}
    \caption{A causal diamond in MSS for a ball-shaped spacelike surface $\Sigma$ with $\partial\Sigma$ as its boundary. $p$ and $p'$ are the future and past vertices of the diamond. The solid curves are the flow lines of the conformal Killing vector $\zeta$. The null boundary $\mathcal{H}$ is the conformal Killing horizon.}
    \label{causal_diamond}
\end{figure}  
\\ A key quantity for our analysis is the Quantum expansion \cite{bousso2016quantum}. Suppose a spatial surface $\sigma$ of area $A$ splits a Cauchy surface $\Sigma$ into two parts. To define quantum expansion($\Theta$) at $\sigma$, erect an orthogonal null hypersurface $N$ and consider the response of $S_{gen}$ to deformations of $\sigma$ along $N$. Furthermore, $N$ can be divided into pencils of width $\mathcal{A}$ around its null generator and $\sigma$ is deformed an affine parameter length $\lambda>0$ along its generator, then $\Theta$ is defined like classical expansion $\theta$ as
\begin{equation}
    \Theta=\lim_{\mathcal{A}\to0}\frac{4G}{\mathcal{A}}\frac{dS_{gen}}{d\lambda}.\label{quantum_expansion}
\end{equation}
This gives the change in generalized entropy as
\begin{equation}
    \delta S_{gen}=\frac{1}{4G}\int_\mathcal{H}\Theta d\lambda d\mathcal{A}.\label{sgen_variation}
\end{equation}
For Einstein's gravity, we only consider the leading contribution coming from the area resulting in 
\begin{equation}
    S_{gen}=\frac{A}{4G}+S_{out}.
\end{equation}
Therefore,
\begin{equation}
\frac{4G}{\mathcal{A}}\frac{dS_{gen}}{d\lambda}=\frac{1}{\mathcal{A}}\frac{dA}{d\lambda}+\frac{4G}{\mathcal{A}}\frac{dS_{out}}{d\lambda}.
\end{equation}
From this, we obtain
\begin{equation}
    \Theta=\theta+\frac{4G}{\mathcal{A}}S'_{out}\hspace{2mm}\text{(Einstein's gravity)},
\end{equation}
where $'$ denotes derivative with respect to the affine parameter $\lambda$ and $\theta$ is the classical expansion defined as
\begin{equation}
    \theta=\frac{1}{\mathcal{A}}\frac{dA}{d\lambda}.
\end{equation}
However, subleading divergences also need to be included for higher curvature gravity. Therefore, the area-entropy relation is replaced with Wald entropy in $S_{gen}$, resulting in
\begin{equation}
      \Theta=\theta_{HC}+\frac{4G}{\mathcal{A}}S'_{out}\hspace{2mm}\text{(General theory of gravity)},
\end{equation}
where a "higher curvature" expansion is defined as
\begin{equation}
    \theta_{HC}=\frac{4G}{\mathcal{A}}\frac{dS_{Wald}}{d\lambda}.
\end{equation} 
The evolution of higher curvature expansion ($\theta_{HC}$) shall be governed by a higher curvature Raychaudhuri equation (see also \cite{wall2024linearized,yan2024gravitationalfocusinghorizonentropy} where similar generalized Raychaudhuri equation is discussed useful for proving focusing theorem in a general theory of gravity)
\begin{equation}
    \frac{d\theta_{HC}}{d\lambda}=-\frac{\theta^2_{HC}}{2}-\xi^2-E_{ab}k^ak^b.\label{modified_raychaudhuri}
\end{equation}
where $E_{ab}$ is given by (\ref{Eab}). For Einstein's gravity, $S_{Wald}=\frac{A}{4G}$, $E_{ab}k^ak^b=R_{ab}k^ak^b$ and therefore, $\theta_{HC}=\theta$. It should be noted that while the Raychaudhuri equation is a kinematical relation valid independently of the theory of gravity, the redefinition of the expansion scalar \( \theta_{\text{HC}} \) to include higher curvature effects modifies the physical interpretation of terms on its right-hand side. In particular, the Ricci curvature term \( R_{ab} k^a k^b \) is replaced by \( E_{ab} k^a k^b \), where \( E_{ab} \) incorporates contributions from the higher curvature corrections. This adjustment ensures consistency with the underlying theory of gravity while preserving the geometric essence of the equation. That's why we explicitly refer to it as the "higher curvature" Raychaudhuri equation. To better understand what $E_{ab}$ is, we write the action as
\begin{equation}
    I=\int d^4x\sqrt{-g}L(g_{ab},R_{abcd},...)+L_m.
\end{equation}
We can make it even more general by adding terms like $\nabla_a R_{bcde}$, $\nabla_a\nabla_b R_{cdef}$ and so on. So, $I$ is the most general action. Varying the action with respect to metric gives the equation of motion as
\begin{equation}
    E_{ab}=\frac{2}{\sqrt{-g}}\frac{\delta L}{\delta g^{ab}}=\frac{-2}{\sqrt{-g}}\frac{\delta L_m}{\delta g^{ab}}\equiv T_{ab}.\label{Eab}
\end{equation}
To derive the semiclassical equation for higher curvature gravity, we first write the infinitesimal evolution of quantum expansion $\Theta$ as we move away from the equilibrium surface, along the null congruence around its equilibrium value at $p$ as
\begin{equation}
    \Theta=\Theta_p+\lambda\frac{d\Theta}{d\lambda}\bigg|_p+\mathcal{O}(\lambda^2).
\end{equation}
Using (\ref{sgen_variation}), the change in generalized entropy is
\begin{equation}
\begin{split}
\delta S_{gen}&=\frac{1}{4G}\int_\mathcal{H}d\lambda d\mathcal{A}\bigg[\Theta_p-\lambda \bigg(\frac{\theta_{HC}^2}{2}+\xi^2+E_{ab}k^ak^b\\
&-\frac{4G}{\mathcal{A}}(S''_{out}-S'_{out}\theta)\bigg)\bigg]_p.   
\end{split}
\end{equation}
The classical expansion $\theta$ of the null boundary for a causal diamond (which is a conformal Killing horizon) does not vanish, however, the shear $\xi$ vanishes \cite{dyer1979conformal}. Thus, it may seem that the notion of dynamical equilibrium (in the sense of physical process) is not well-defined. However, there is a way ahead: To impose the dynamic equilibrium condition $\delta S_{\text{gen}} = 0$, we use the restricted quantum focusing \cite{shahbazi2024restricted}, which allows the quantum expansion $\Theta$ to vanish pointwise (restricted quantum focusing is a weaker constraint than quantum focusing and has been proven in Braneworld scenario, making it a robust result). This ensures that the generalized entropy does not change along the null generator on the boundary $\mathcal{H}$ of the diamond. This means that to obtain the equation of motion, we require the null boundary $\mathcal{H}$ to be extremal \cite{brown2020python,akers2020quantum,engelhardt2021world}. This is also supported by AdS/CFT \cite{maldacena1999large}, which states that any CFT subsystem is dual to a quantum extremal region \cite{engelhardt2015quantum,dong2018entropy}, i.e., the generalized entropy is stationary. Moving further in the discussion, even if the classical expansion \( \theta \) does not vanish, we can argue that \( \theta \) (or $\theta_{HC}$) is small due to the requirement of equilibrium conditions. Since the quantum expansion \( \Theta \) vanishes, \( S'_{\text{out}} \) must also be small. As a result, the terms \( \theta_{HC}^2 \) and \( S'_{\text{out}} \theta \) are higher order and can be safely ignored in this context. This gives the change in generalized entropy as
\begin{equation}
    \delta S_{gen}=-\frac{1}{4G}\int_\mathcal{H}d\lambda d\mathcal{A}\lambda\bigg(E_{kk}-\frac{4G}{\mathcal{A}}S''_{out}\bigg).
\end{equation}
Here, $E_{kk}$ denotes the null-null component of $E_{ab}$.
For CFTs, we have the important result \cite{leichenauer2018energy,balakrishnan2022entropy} 
\begin{equation}
    S''_{out}=2\pi\mathcal{A}\langle T_{kk}\rangle.
\end{equation}
We assume that matter is described by a QFT with a UV fixed point and that the length of the diamond is much smaller than the QFT scale but larger than the Planck scale\footnote{This is also the assumption of Jacobson's entanglement equilibrium approach.} ($l_{p}<<l<<l_{QFT}$). Then, for the small diamond, we can map the CFT result above to any QFT. This gives
\begin{equation}
    \delta S_{gen}=-\frac{1}{4G}\int_\mathcal{H}d\lambda d\mathcal{A}\lambda\bigg(E_{kk}-8\pi G\langle T_{kk}\rangle\bigg).\label{delta_sgen_zero}
\end{equation}
We finally demand the equilibrium condition
\begin{equation}
    \delta S_{gen}=0.
\end{equation}
Since the integrand vanishes for all null vectors pointwise (a local constraint), we obtain the full non-linear semiclassical equation of motion valid at all points in spacetime as
\begin{equation}
    E_{ab}+\Lambda g_{ab}=8\pi G\langle T_{ab}\rangle.
\end{equation}
with an undetermined constant $\Lambda$ identified as the cosmological constant as usual.
\section{Conclusion and Discussion}
In this letter, we have successfully derived the full non-linear semiclassical equation of motion for a general diffeomorphism invariant theory of gravity. We presented two approaches: one perturbative quantum gravity approach applied to a Rindler horizon and mapped to a stretched future light cone (a spherical Rindler horizon) and another using the "higher curvature" Raychaudhuri equation applied to a causal diamond in MSS. Earlier works \cite{bueno2017entanglement,svesko2019entanglement} derived only the linearized equations using Jacobson's entanglement equilibrium approach \cite{jacobson2016entanglement} and concluded that non-linear equations cannot be derived using this approach. This remained an open problem since. Instead of entanglement equilibrium, which is a variational approach, we made use of a semiclassical physical process version given as \cite{kumar2023recovering}
\begin{equation}
   \langle Q\rangle=T\delta S_{gen}.
\end{equation} 
The variational version is an equilibrium approach, while the semiclassical physical process version is reminiscent of non-equilibrium thermodynamics. The essence of the semiclassical physical process version $\langle Q\rangle=T\delta S_{gen}$ is that in the semiclassical case, horizon generators do not always focus to the past due to the presence of quantum fields, which violates the classical focusing and thus $\delta A$ can decrease. Therefore, identifying it with the entropy of the system in the Clausius relation is not justified in the semiclassical case. Moreover, classical expansion $\theta$ does not vanish in the semiclassical case, resulting from the failure of classical focusing to hold, making it impossible to implement local equilibrium conditions. However, the generalized entropy always increases, even in the semiclassical case. Therefore, we should identify horizon and matter as a "system", and this generalized entropy should enter the Clausius relation $Q=T\delta S$.\\ A quantum focusing conjecture proposed in \cite{bousso2016quantum} states that the quantum expansion($\Theta$) decreases in the semiclassical case, and thus, a quantum focusing holds. Therefore, at the zeroth order, $\Theta$ can be always chosen to vanish. This imposes a condition on the spacetime curvature, and we can recover Einstein's equation as shown in \cite{kumar2023recovering}. In this work, we proposed an equilibrium condition that corresponds to the vanishing of quantum expansion $\Theta$ and naturally follows from restricted quantum focusing \cite{shahbazi2024restricted}
\begin{equation}
    \delta S_{gen}=0.
\end{equation}
This condition, along with $S''_{out}=2\pi \langle T_{kk}\rangle$ to derive the non-linear equation of motion. We used this approach in section 3. Jacobson's entanglement equilibrium approach proposed a maximum vacuum entanglement hypothesis (MVEH) so that variation in entanglement entropy vanishes for a small ball. In our case, the vanishing of the generalized entropy is a direct result of restricted quantum focusing (a proven result) and is, therefore, robust.\\ One can broadly divide the ways of deriving Einstein's equation into three parts: spacetime thermodynamics \cite{jacobson1995thermodynamics,parikh2016generalized,parikh2018einstein,kumar2023recovering}, spacetime entanglement \cite{lashkari2014gravitational,faulkner2014gravitation,swingle2014universality} and spacetime complexity \cite{pedraza2021lorentzian,pedraza2022sewing,carrasco2023gravitation}. In the spacetime thermodynamics approach, a notion of entropy is assigned to a local causal horizon, and the Clausius relation \( Q = T \delta S_{\text{rev}} \), where \( S_{\text{rev}} \) denotes the reversible part of the entropy, is utilized to derive Einstein's equations and higher curvature equations of motion. In contrast, the spacetime entanglement approach employs the first law of entanglement entropy, \( \delta S_{\text{EE}} = \delta \langle H \rangle \), where \( H \) is the modular Hamiltonian, and then applies the Ryu-Takayanagi (RT) formula \cite{ryu2006aspects,ryu2006holographic} to derive the equations of motion. Jacobson's entanglement equilibrium approach is distinct, as it proceeds directly within spacetime without relying on holography, but it remains conceptually related to the spacetime entanglement paradigm. Our approach aligns closely with the spacetime entanglement framework, although the latter primarily derives the linearized equations of motion. Finally, Einstein's equations can also be derived using the first law of spacetime complexity, expressed as  
\[
\delta \mathcal{C} = (\dot{\lambda}^a|_{\lambda_f}) \eta_{ab} \delta \lambda^b_f,
\]
where \( \mathcal{C} \) represents a specific notion of boundary complexity that quantifies the minimal number of sources \( \{ \lambda_a \} \) required to prepare a holographic CFT state via a Euclidean path integral, and \( \eta_{ab} \) is the metric in the auxiliary space of these sources.
\\In conclusion, this work not only derives the non-linear equation of motion for a general diffeomorphism-invariant theory of gravity but also demonstrates the limitations of the variational approach. While the variational approach is a robust technique, it fails to reveal the full non-linear structure of the equation of motion for higher curvature gravity. In contrast, the physical process approach preserves this structure, proving more effective in capturing the complete dynamics. By utilizing generalized entropy, a cutoff-independent quantity, this work provides insights into the full quantum theory of gravity. While Jacobson’s entanglement equilibrium approach also uses the total entanglement entropy of the diamond—another cutoff-independent quantity (assuming the quantum theory of gravity is UV finite so that the cutoff $\epsilon\sim l_p$)—it cannot be extended to include non-linear corrections in higher curvature gravity. The higher-order non-linear corrections derived here in semiclassical gravity, based on thermodynamic arguments, strengthen the role of higher curvature corrections in any complete theory of (quantum) gravity, which are typically thought to arise from loop quantum gravity or string theory. Furthermore, this work highlights a strong connection between quantum focusing, extremal surface and semiclassical gravity, suggesting that further exploration of this relationship could yield valuable insights. An exciting direction for future work is to explore potential connections between this work (or the broader field of spacetime thermodynamics) and the spacetime complexity program. We are hopeful that these results will inspire future studies and lead to a deeper understanding of quantum gravity.
\section*{Conflict of Interest}
The author declares no conflict of interest.
\section*{Acknowledgements}
It is a pleasure to thank the anonymous referee whose detailed and constructive comments greatly improved the original draft. I am also thankful to Sreejith Nair, Chanchal Sharma and Soham Acharyya for some helpful discussions.
\bibliography{bib}

\begin{thebibliography}{40}
\expandafter\ifx\csname natexlab\endcsname\relax\def\natexlab#1{#1}\fi
\providecommand{\url}[1]{\texttt{#1}}
\providecommand{\href}[2]{#2}
\providecommand{\path}[1]{#1}
\providecommand{\DOIprefix}{doi:}
\providecommand{\ArXivprefix}{arXiv:}
\providecommand{\URLprefix}{URL: }
\providecommand{\Pubmedprefix}{pmid:}
\providecommand{\doi}[1]{\href{http://dx.doi.org/#1}{\path{#1}}}
\providecommand{\Pubmed}[1]{\href{pmid:#1}{\path{#1}}}
\providecommand{\bibinfo}[2]{#2}
\ifx\xfnm\relax \def\xfnm[#1]{\unskip,\space#1}\fi
\bibitem[{Akers et~al.(2020)Akers, Engelhardt, Penington and Usatyuk}]{akers2020quantum}
\bibinfo{author}{Akers, C.}, \bibinfo{author}{Engelhardt, N.}, \bibinfo{author}{Penington, G.}, \bibinfo{author}{Usatyuk, M.}, \bibinfo{year}{2020}.
\newblock \bibinfo{title}{Quantum maximin surfaces}.
\newblock \bibinfo{journal}{Journal of High Energy Physics} \bibinfo{volume}{2020}, \bibinfo{pages}{1--43}.
\bibitem[{Balakrishnan et~al.(2022)Balakrishnan, Chandrasekaran, Faulkner, Levine and Shahbazi-Moghaddam}]{balakrishnan2022entropy}
\bibinfo{author}{Balakrishnan, S.}, \bibinfo{author}{Chandrasekaran, V.}, \bibinfo{author}{Faulkner, T.}, \bibinfo{author}{Levine, A.}, \bibinfo{author}{Shahbazi-Moghaddam, A.}, \bibinfo{year}{2022}.
\newblock \bibinfo{title}{Entropy variations and light ray operators from replica defects}.
\newblock \bibinfo{journal}{Journal of High Energy Physics} \bibinfo{volume}{2022}, \bibinfo{pages}{1--42}.
\bibitem[{Bekenstein(1972)}]{Bekenstein:1972tm}
\bibinfo{author}{Bekenstein, J.D.}, \bibinfo{year}{1972}.
\newblock \bibinfo{title}{{Black holes and the second law}}.
\newblock \bibinfo{journal}{Lett. Nuovo Cim.} \bibinfo{volume}{4}, \bibinfo{pages}{737--740}.
\newblock \DOIprefix\doi{10.1007/BF02757029}.
\bibitem[{Bekenstein(1973)}]{bekenstein1973black}
\bibinfo{author}{Bekenstein, J.D.}, \bibinfo{year}{1973}.
\newblock \bibinfo{title}{Black holes and entropy}.
\newblock \bibinfo{journal}{Physical Review D} \bibinfo{volume}{7}, \bibinfo{pages}{2333}.
\bibitem[{Bianchi(2013)}]{bianchi2013blackholeentropygraviton}
\bibinfo{author}{Bianchi, E.}, \bibinfo{year}{2013}.
\newblock \bibinfo{title}{Black hole entropy from graviton entanglement}.
\newblock \URLprefix \url{https://arxiv.org/abs/1211.0522}, \href{http://arxiv.org/abs/1211.0522}{{\tt arXiv:1211.0522}}.
\bibitem[{Bianchi and Satz(2013)}]{bianchi2013mechanical}
\bibinfo{author}{Bianchi, E.}, \bibinfo{author}{Satz, A.}, \bibinfo{year}{2013}.
\newblock \bibinfo{title}{Mechanical laws of the rindler horizon}.
\newblock \bibinfo{journal}{Physical Review D} \bibinfo{volume}{87}, \bibinfo{pages}{124031}.
\bibitem[{Bousso et~al.(2016)Bousso, Fisher, Leichenauer and Wall}]{bousso2016quantum}
\bibinfo{author}{Bousso, R.}, \bibinfo{author}{Fisher, Z.}, \bibinfo{author}{Leichenauer, S.}, \bibinfo{author}{Wall, A.C.}, \bibinfo{year}{2016}.
\newblock \bibinfo{title}{Quantum focusing conjecture}.
\newblock \bibinfo{journal}{Physical Review D} \bibinfo{volume}{93}, \bibinfo{pages}{064044}.
\bibitem[{Brown et~al.(2020)Brown, Gharibyan, Penington and Susskind}]{brown2020python}
\bibinfo{author}{Brown, A.R.}, \bibinfo{author}{Gharibyan, H.}, \bibinfo{author}{Penington, G.}, \bibinfo{author}{Susskind, L.}, \bibinfo{year}{2020}.
\newblock \bibinfo{title}{The python’s lunch: geometric obstructions to decoding hawking radiation}.
\newblock \bibinfo{journal}{Journal of High Energy Physics} \bibinfo{volume}{2020}, \bibinfo{pages}{1--53}.
\bibitem[{Bueno et~al.(2017)Bueno, Min, Speranza and Visser}]{bueno2017entanglement}
\bibinfo{author}{Bueno, P.}, \bibinfo{author}{Min, V.S.}, \bibinfo{author}{Speranza, A.J.}, \bibinfo{author}{Visser, M.R.}, \bibinfo{year}{2017}.
\newblock \bibinfo{title}{Entanglement equilibrium for higher order gravity}.
\newblock \bibinfo{journal}{Physical Review D} \bibinfo{volume}{95}, \bibinfo{pages}{046003}.
\bibitem[{Carrasco et~al.(2023)Carrasco, Pedraza, Svesko and Weller-Davies}]{carrasco2023gravitation}
\bibinfo{author}{Carrasco, R.}, \bibinfo{author}{Pedraza, J.F.}, \bibinfo{author}{Svesko, A.}, \bibinfo{author}{Weller-Davies, Z.}, \bibinfo{year}{2023}.
\newblock \bibinfo{title}{Gravitation from optimized computation: Einstein and beyond}.
\newblock \bibinfo{journal}{Journal of High Energy Physics} \bibinfo{volume}{2023}, \bibinfo{pages}{1--48}.
\bibitem[{Dong and Lewkowycz(2018)}]{dong2018entropy}
\bibinfo{author}{Dong, X.}, \bibinfo{author}{Lewkowycz, A.}, \bibinfo{year}{2018}.
\newblock \bibinfo{title}{Entropy, extremality, euclidean variations, and the equations of motion}.
\newblock \bibinfo{journal}{Journal of High Energy Physics} \bibinfo{volume}{2018}, \bibinfo{pages}{1--33}.
\bibitem[{Dyer and Honig(1979)}]{dyer1979conformal}
\bibinfo{author}{Dyer, C.}, \bibinfo{author}{Honig, E.}, \bibinfo{year}{1979}.
\newblock \bibinfo{title}{Conformal killing horizons}.
\newblock \bibinfo{journal}{Journal of Mathematical Physics} \bibinfo{volume}{20}, \bibinfo{pages}{409--412}.
\bibitem[{Eling et~al.(2006)Eling, Guedens and Jacobson}]{eling2006nonequilibrium}
\bibinfo{author}{Eling, C.}, \bibinfo{author}{Guedens, R.}, \bibinfo{author}{Jacobson, T.}, \bibinfo{year}{2006}.
\newblock \bibinfo{title}{Nonequilibrium thermodynamics of spacetime}.
\newblock \bibinfo{journal}{Physical Review Letters} \bibinfo{volume}{96}, \bibinfo{pages}{121301}.
\bibitem[{Engelhardt et~al.(2021)Engelhardt, Penington and Shahbazi-Moghaddam}]{engelhardt2021world}
\bibinfo{author}{Engelhardt, N.}, \bibinfo{author}{Penington, G.}, \bibinfo{author}{Shahbazi-Moghaddam, A.}, \bibinfo{year}{2021}.
\newblock \bibinfo{title}{A world without pythons would be so simple}.
\newblock \bibinfo{journal}{Classical and Quantum Gravity} \bibinfo{volume}{38}, \bibinfo{pages}{234001}.
\bibitem[{Engelhardt and Wall(2015)}]{engelhardt2015quantum}
\bibinfo{author}{Engelhardt, N.}, \bibinfo{author}{Wall, A.C.}, \bibinfo{year}{2015}.
\newblock \bibinfo{title}{Quantum extremal surfaces: holographic entanglement entropy beyond the classical regime}.
\newblock \bibinfo{journal}{Journal of High Energy Physics} \bibinfo{volume}{2015}, \bibinfo{pages}{1--27}.
\bibitem[{Faulkner et~al.(2014)Faulkner, Guica, Hartman, Myers and Van~Raamsdonk}]{faulkner2014gravitation}
\bibinfo{author}{Faulkner, T.}, \bibinfo{author}{Guica, M.}, \bibinfo{author}{Hartman, T.}, \bibinfo{author}{Myers, R.C.}, \bibinfo{author}{Van~Raamsdonk, M.}, \bibinfo{year}{2014}.
\newblock \bibinfo{title}{Gravitation from entanglement in holographic cfts}.
\newblock \bibinfo{journal}{Journal of High Energy Physics} \bibinfo{volume}{2014}, \bibinfo{pages}{1--41}.
\bibitem[{Guedens et~al.(2012)Guedens, Jacobson and Sarkar}]{guedens2012horizon}
\bibinfo{author}{Guedens, R.}, \bibinfo{author}{Jacobson, T.}, \bibinfo{author}{Sarkar, S.}, \bibinfo{year}{2012}.
\newblock \bibinfo{title}{Horizon entropy and higher curvature equations of state}.
\newblock \bibinfo{journal}{Physical Review D—Particles, Fields, Gravitation, and Cosmology} \bibinfo{volume}{85}, \bibinfo{pages}{064017}.
\bibitem[{Hawking(1975)}]{hawking1975particle}
\bibinfo{author}{Hawking, S.W.}, \bibinfo{year}{1975}.
\newblock \bibinfo{title}{Particle creation by black holes}.
\newblock \bibinfo{journal}{Communications in mathematical physics} \bibinfo{volume}{43}, \bibinfo{pages}{199--220}.
\bibitem[{Jacobson(1995)}]{jacobson1995thermodynamics}
\bibinfo{author}{Jacobson, T.}, \bibinfo{year}{1995}.
\newblock \bibinfo{title}{Thermodynamics of spacetime: the einstein equation of state}.
\newblock \bibinfo{journal}{Physical Review Letters} \bibinfo{volume}{75}, \bibinfo{pages}{1260}.
\bibitem[{Jacobson(2016)}]{jacobson2016entanglement}
\bibinfo{author}{Jacobson, T.}, \bibinfo{year}{2016}.
\newblock \bibinfo{title}{Entanglement equilibrium and the einstein equation}.
\newblock \bibinfo{journal}{Physical review letters} \bibinfo{volume}{116}, \bibinfo{pages}{201101}.
\bibitem[{Kumar(2023)}]{kumar2023recovering}
\bibinfo{author}{Kumar, N.}, \bibinfo{year}{2023}.
\newblock \bibinfo{title}{Recovering semiclassical einstein’s equation using generalized entropy}.
\newblock \bibinfo{journal}{General Relativity and Gravitation} \bibinfo{volume}{55}, \bibinfo{pages}{127}.
\bibitem[{Lashkari et~al.(2014)Lashkari, McDermott and Van~Raamsdonk}]{lashkari2014gravitational}
\bibinfo{author}{Lashkari, N.}, \bibinfo{author}{McDermott, M.B.}, \bibinfo{author}{Van~Raamsdonk, M.}, \bibinfo{year}{2014}.
\newblock \bibinfo{title}{Gravitational dynamics from entanglement “thermodynamics”}.
\newblock \bibinfo{journal}{Journal of High Energy Physics} \bibinfo{volume}{2014}, \bibinfo{pages}{1--16}.
\bibitem[{Leichenauer et~al.(2018)Leichenauer, Levine and Shahbazi-Moghaddam}]{leichenauer2018energy}
\bibinfo{author}{Leichenauer, S.}, \bibinfo{author}{Levine, A.}, \bibinfo{author}{Shahbazi-Moghaddam, A.}, \bibinfo{year}{2018}.
\newblock \bibinfo{title}{Energy density from second shape variations of the von neumann entropy}.
\newblock \bibinfo{journal}{Physical Review D} \bibinfo{volume}{98}, \bibinfo{pages}{086013}.
\bibitem[{Maldacena(1999)}]{maldacena1999large}
\bibinfo{author}{Maldacena, J.}, \bibinfo{year}{1999}.
\newblock \bibinfo{title}{The large-n limit of superconformal field theories and supergravity}.
\newblock \bibinfo{journal}{International journal of theoretical physics} \bibinfo{volume}{38}, \bibinfo{pages}{1113--1133}.
\bibitem[{Parikh and Sarkar(2016)}]{parikh2016generalized}
\bibinfo{author}{Parikh, M.}, \bibinfo{author}{Sarkar, S.}, \bibinfo{year}{2016}.
\newblock \bibinfo{title}{Generalized einstein’s equations from wald entropy}.
\newblock \bibinfo{journal}{Entropy} \bibinfo{volume}{18}, \bibinfo{pages}{119}.
\bibitem[{Parikh et~al.(2020)Parikh, Sarkar and Svesko}]{parikh2020local}
\bibinfo{author}{Parikh, M.}, \bibinfo{author}{Sarkar, S.}, \bibinfo{author}{Svesko, A.}, \bibinfo{year}{2020}.
\newblock \bibinfo{title}{Local first law of gravity}.
\newblock \bibinfo{journal}{Physical Review D} \bibinfo{volume}{101}, \bibinfo{pages}{104043}.
\bibitem[{Parikh and Svesko(2018)}]{parikh2018einstein}
\bibinfo{author}{Parikh, M.}, \bibinfo{author}{Svesko, A.}, \bibinfo{year}{2018}.
\newblock \bibinfo{title}{Einstein’s equations from the stretched future light cone}.
\newblock \bibinfo{journal}{Physical Review D} \bibinfo{volume}{98}, \bibinfo{pages}{026018}.
\bibitem[{Pedraza et~al.(2021)Pedraza, Russo, Svesko and Weller-Davies}]{pedraza2021lorentzian}
\bibinfo{author}{Pedraza, J.F.}, \bibinfo{author}{Russo, A.}, \bibinfo{author}{Svesko, A.}, \bibinfo{author}{Weller-Davies, Z.}, \bibinfo{year}{2021}.
\newblock \bibinfo{title}{Lorentzian threads as gatelines and holographic complexity}.
\newblock \bibinfo{journal}{Physical Review Letters} \bibinfo{volume}{127}, \bibinfo{pages}{271602}.
\bibitem[{Pedraza et~al.(2022)Pedraza, Russo, Svesko and Weller-Davies}]{pedraza2022sewing}
\bibinfo{author}{Pedraza, J.F.}, \bibinfo{author}{Russo, A.}, \bibinfo{author}{Svesko, A.}, \bibinfo{author}{Weller-Davies, Z.}, \bibinfo{year}{2022}.
\newblock \bibinfo{title}{Sewing spacetime with lorentzian threads: complexity and the emergence of time in quantum gravity}.
\newblock \bibinfo{journal}{Journal of High Energy Physics} \bibinfo{volume}{2022}, \bibinfo{pages}{1--123}.
\bibitem[{Ryu and Takayanagi(2006a)}]{ryu2006aspects}
\bibinfo{author}{Ryu, S.}, \bibinfo{author}{Takayanagi, T.}, \bibinfo{year}{2006}a.
\newblock \bibinfo{title}{Aspects of holographic entanglement entropy}.
\newblock \bibinfo{journal}{Journal of High Energy Physics} \bibinfo{volume}{2006}, \bibinfo{pages}{045}.
\bibitem[{Ryu and Takayanagi(2006b)}]{ryu2006holographic}
\bibinfo{author}{Ryu, S.}, \bibinfo{author}{Takayanagi, T.}, \bibinfo{year}{2006}b.
\newblock \bibinfo{title}{Holographic derivation of entanglement entropy from the anti--de sitter space/conformal field theory correspondence}.
\newblock \bibinfo{journal}{Physical review letters} \bibinfo{volume}{96}, \bibinfo{pages}{181602}.
\bibitem[{Shahbazi-Moghaddam(2024)}]{shahbazi2024restricted}
\bibinfo{author}{Shahbazi-Moghaddam, A.}, \bibinfo{year}{2024}.
\newblock \bibinfo{title}{Restricted quantum focusing}.
\newblock \bibinfo{journal}{Physical Review D} \bibinfo{volume}{109}, \bibinfo{pages}{066023}.
\bibitem[{Solodukhin(2011)}]{solodukhin2011entanglement}
\bibinfo{author}{Solodukhin, S.N.}, \bibinfo{year}{2011}.
\newblock \bibinfo{title}{Entanglement entropy of black holes}.
\newblock \bibinfo{journal}{Living Reviews in Relativity} \bibinfo{volume}{14}, \bibinfo{pages}{1--96}.
\bibitem[{Sorkin(1986)}]{sorkin1986toward}
\bibinfo{author}{Sorkin, R.D.}, \bibinfo{year}{1986}.
\newblock \bibinfo{title}{Toward a proof of entropy increase in the presence of quantum black holes}.
\newblock \bibinfo{journal}{Physical review letters} \bibinfo{volume}{56}, \bibinfo{pages}{1885}.
\bibitem[{Sorkin(1997)}]{sorkin1997statisticalmechanicsblackhole}
\bibinfo{author}{Sorkin, R.D.}, \bibinfo{year}{1997}.
\newblock \bibinfo{title}{The statistical mechanics of black hole thermodynamics}.
\newblock \URLprefix \url{https://arxiv.org/abs/gr-qc/9705006}, \href{http://arxiv.org/abs/gr-qc/9705006}{{\tt arXiv:gr-qc/9705006}}.
\bibitem[{Susskind and Uglum(1994)}]{susskind1994black}
\bibinfo{author}{Susskind, L.}, \bibinfo{author}{Uglum, J.}, \bibinfo{year}{1994}.
\newblock \bibinfo{title}{Black hole entropy in canonical quantum gravity and superstring theory}.
\newblock \bibinfo{journal}{Physical Review D} \bibinfo{volume}{50}, \bibinfo{pages}{2700}.
\bibitem[{Svesko(2019)}]{svesko2019entanglement}
\bibinfo{author}{Svesko, A.}, \bibinfo{year}{2019}.
\newblock \bibinfo{title}{From entanglement to thermodynamics and to gravity}.
\newblock \bibinfo{journal}{Physical Review D} \bibinfo{volume}{99}, \bibinfo{pages}{086006}.
\bibitem[{Swingle and Van~Raamsdonk(2014)}]{swingle2014universality}
\bibinfo{author}{Swingle, B.}, \bibinfo{author}{Van~Raamsdonk, M.}, \bibinfo{year}{2014}.
\newblock \bibinfo{title}{Universality of gravity from entanglement}.
\newblock \bibinfo{journal}{arXiv preprint arXiv:1405.2933} .
\bibitem[{Wall and Yan(2024)}]{wall2024linearized}
\bibinfo{author}{Wall, A.C.}, \bibinfo{author}{Yan, Z.}, \bibinfo{year}{2024}.
\newblock \bibinfo{title}{Linearized second law for higher curvature gravity and nonminimally coupled vector fields}.
\newblock \bibinfo{journal}{Physical Review D} \bibinfo{volume}{110}, \bibinfo{pages}{084005}.
\bibitem[{Yan(2024)}]{yan2024gravitationalfocusinghorizonentropy}
\bibinfo{author}{Yan, Z.}, \bibinfo{year}{2024}.
\newblock \bibinfo{title}{Gravitational focusing and horizon entropy for higher-spin fields}.
\newblock \URLprefix \url{https://arxiv.org/abs/2412.07107}, \href{http://arxiv.org/abs/2412.07107}{{\tt arXiv:2412.07107}}.

\end{thebibliography}
\bibliographystyle{elsarticle-harv}

\end{document}